\documentclass[lettersize,journal]{IEEEtran}
\usepackage{amsmath,amsfonts}
\usepackage{algorithmic}
\usepackage{algorithm}
\usepackage{array}
\usepackage[caption=false,font=normalsize,labelfont=sf,textfont=sf]{subfig}
\usepackage{textcomp}
\usepackage{stfloats}
\usepackage{url}
\usepackage{verbatim}
\usepackage{graphicx}
\usepackage{cite}

\begin{document}

\title{Low-Altitude Agentic Networks for Optical Wireless Communication \\and Sensing: An Oceanic Scenario}

\author{Tianqi Mao,~\IEEEmembership{Member,~IEEE}, Jiayue Liu,~\IEEEmembership{Student Member,~IEEE}, Zeping Sui,~\IEEEmembership{Member,~IEEE}, Leyu Cao, \\Xiao Liang, Dezhi Zheng, and Zhaocheng Wang,~\IEEEmembership{Fellow,~IEEE}
\thanks{T. Mao, J. Liu, L. Cao, X. Liang and D. Zheng are with the State Key Laboratory of Environment Characteristics and Effects for Near-space, Beijing Institute of Technology, Beijing 100081, China (e-mails: maotq@bit.edu.cn, jiayue\_liu@bit.edu.cn, leyu\_cao@bit.edu.cn, liang.xiao@bit.edu.cn, zhengdezhi@bit.edu.cn).}
\thanks{Z. Sui is with the School of Computer Science and
Electronics Engineering, University of Essex, CO4 3SQ Colchester, U.K.
(e-mail: zepingsui@outlook.com).}
\thanks{Z.~Wang is with Department of Electronic Engineering, Tsinghua University, Beijing 100084, China (e-mail:~zcwang@tsinghua.edu.cn).}

}



\maketitle

\begin{abstract}

The cross-domain oceanic connectivity ranging from underwater to the sky has become increasingly indispensable for a plethora of data-consuming maritime applications, such as maritime meteorological monitoring and offshore exploration. However, broadband implementations can be severely hindered by the isolation from terrestrial networks, limited satellite resources, and the fundamental inability of radio waves to bridge the water-air interface at high rates. 
To this end, this paper introduces an optical network bridging underwater, air and near space, which features a number of cooperative low-altitude platforms (LAPs), serving as compute-capable, sensing-aware, and mission-adaptive agents. 
The network architecture consists of three scenario-specific segments, i.e., water-air direct link, low-altitude mesh network, and the near-space access network. With coordinate sensing and intelligent control, the system tightly couples beam tracking and resource optimization, enabling resilient networking under high mobility and harsh maritime dynamics. 
Furthermore, we review enabling technologies spanning from water-air channel modeling, adaptive beam alignment under sea-surface perturbations, to swarm-intelligence networking for decentralized control, integrated pose-topology planning, and optical Integrated sensing and communication (ISAC) for near-space target detection and beam alignment. 
Finally, open issues are also highlighted, constituting a clear roadmap toward scalable, secure, and ultra-broadband oceanic optical networks.
\end{abstract}


\section{Introduction}
The oceanic activities have become an ever-increasing tendency due to the expanding border of human activities. Various modern open-sea activities, such as maritime meteorological monitoring, maritime target reconnaissance, underwater operation, and emergency communication, necessitate communication and networking access for massive data transmission, precise equipment control, and appropriate device deployment\cite{Survey_SAGS}.
However, the existing architecture with isolated links and limited satellite resources offers limited bandwidth, unstable data rates, and insufficient flexibility\cite{Maritime_com}, making it difficult to satisfy the requirements of the bandwidth-consuming maritime applications with diverse missions, heterogeneous nodes.
To address the challenges, the low altitude wireless network (LAWN) can be introduced as a promising solution to this issue by connecting underwater, aerial, near-space, and terrestrial nodes, and providing flexible access for maritime applications within the near-space-air-ground-sea integrated network architecture.
This oceanic LAWN is mainly constituted by various agentic low-altitude platforms (LAPs) represented by unmanned aerial vehicles (UAVs) with high mobility, distributed network structure, active sensing ability, and on-board intelligence. Hence, it enables fast deployment, intelligent load balancing, and adaptive beam adjustment to support the cooperation of maritime-based and near-space units with lower latencies and higher rate than traditional satellite approaches\cite{NS-ISAC}.

Classical LAWN typically employs full radio-frequency (RF) communications for data exchange \cite{ISAC_LAE}, which, however, suffers from bandwidth limitations and eavesdropping risks. 
Besides, it becomes no longer suitable when communicating with underwater platforms like UUVs or submarines, due to its severe attenuation in water media.
To this end, optical wireless communication (OWC), which possesses stronger penetrability through water and sufficient available bandwidth, has emerged as an innovative scheme for  LAP communication and sensing\cite{Sensing_OWC}. However, there are still technical challenges against the optical LAWN implementations at different communication scenarios, which are highlighted as below:
\begin{enumerate}
    \item {\bf Underwater-Air Direct Link}: The cross-interface transmission is inevitably hindered by the wavy sea surface, causing time-variant deviation of the beam directions, which necessitates real-time channel sensing and dynamic beam alignment schemes to guarantee stable transceiver gain \cite{Underwater_and}.
    \item{\bf Inter-LAP Network}: networking within the LAP swarm suffers from strong Doppler shifts under high mobility from the physical-layer perspective. Besides, sophisticated network-level optimization of the LAWN core structures can be also indispensable \cite{UAV_coop}.
    \item{\bf Near-Space-Air Transmission}: Reliable wireless backhaul faces the peculiar channel nonuniformity and severe pointing errors under extremely-long-range transmission, requiring efficient beam-alignment empowered by emerging technologies like artificial intelligence and integrated sensing and communication (ISAC) \cite{NearspaceTHz}.
    
\end{enumerate}
This article presents an extensive research on the LAWN-centric near-space-air-ground-sea optical network, while systematically analyzing the critical technologies that constitute the system. 
In the following sections, we first introduce the overall system architecture distinguished by links in different scenarios by discussing the software/hardware constitution of each part and analyzing performances of the existing realizations. 
We then analyze the key technologies involved in each part of the network, discussing their roles in increasing link connection, network optimizing, and intelligent augmentation. 
Finally, we study several open challenges and possible research directions that are worthy of exploration, highlighting future directions and possible opportunities for the development of related technologies.

\section{Network Architecture}\label{s2}

\begin{figure*}
    \centering
    \includegraphics[width=1\linewidth]{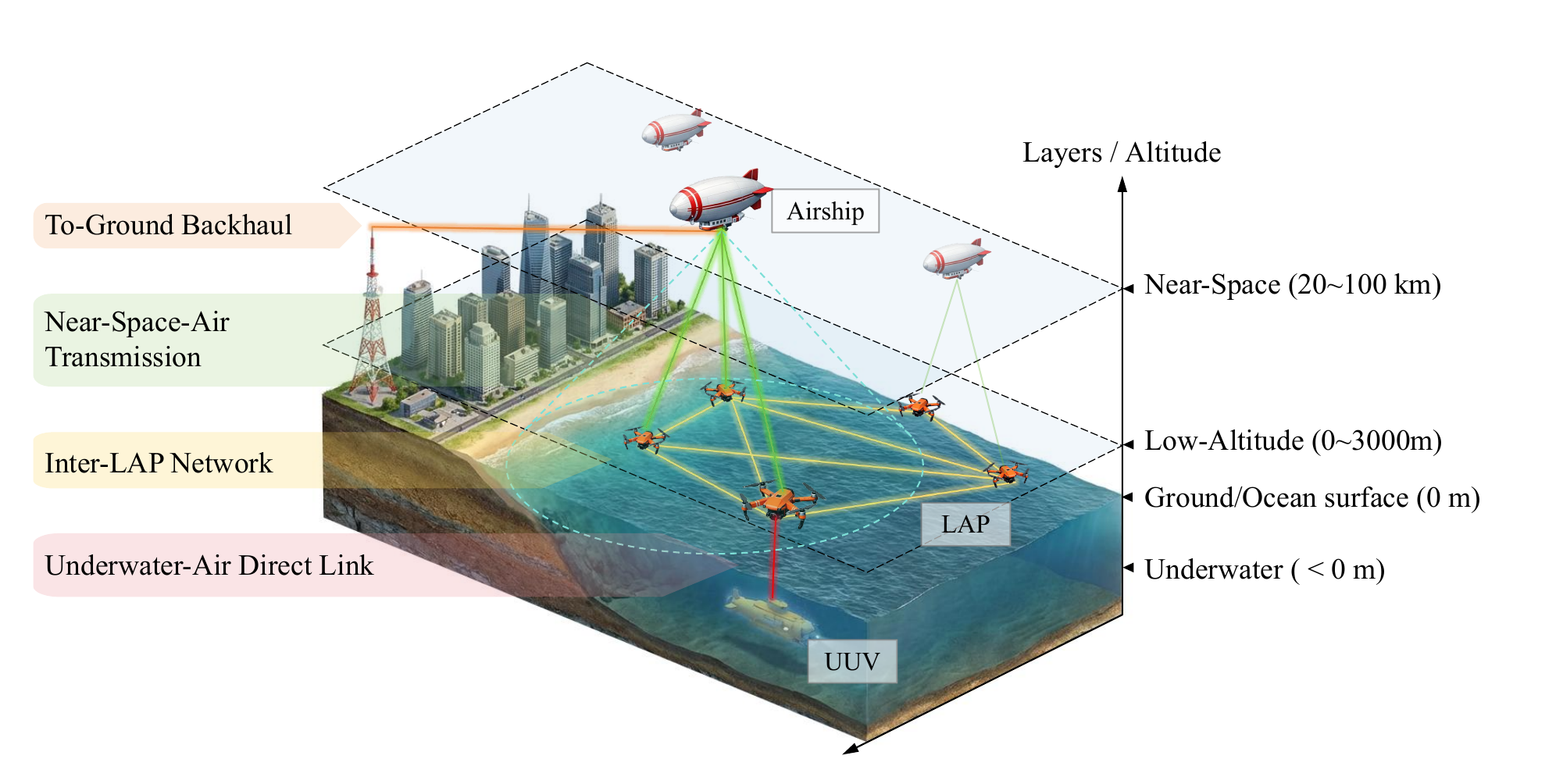}
    \caption{Network architecture of the LAWN-centric near-space-air-ground-sea integrated optical network, which operates in an oceanic scenario.}
    \label{system}
\end{figure*}

\subsection{Network Structure}\label{s2.1}

As shown in Fig. \ref{system}, a LAWN-centric near-space-air-ground-sea optical network is generally implemented as a multilayer architecture in which platforms in different layers interconnect through cross-domain gateways and unified networking management. 
The network architecture can be divided into underwater, ground, low-altitude aerial, and near-space layers based on the physical altitude.
The near-space layer, consisting of near-space platforms such as airships and tethered balloons, provides wide coverage and long-range communication and positioning services, supporting beyond-line-of-sight transport in the ocean-to-land scenario. 
The low-altitude aerial layer, consisting of UAVs and aerostats, functions as deployable networking and sensing nodes, enabling rapid coverage extension and connectivity bridging by mobilizing agentic  LAP nodes. 
The underwater layer, consisting of UUVs, serves as the terminal execution unit that performs oceanic tasks under the network's unified support by maintaining connections with  LAPs. 
The ground layer provides primary compute and storage resources via massive computing clusters, enabling remote intelligent support, centralized data processing, and globalized resource allocation across the network. 
Within the aforementioned architecture, optical wireless communication (OWC) has emerged as an innovative technology for intra-layer and inter-layer connectivity, thereby forming the overall network.
The high transmission rate of OWC carriers naturally addresses the growing need for high-throughput links. Moreover, its superior penetration performance offers a practical approach to establish cross-interface links from underwater to air. 




\subsection{Underwater-Air Direct Optical Link}\label{s2.1.1}
The water-air direct link connects underwater nodes (e.g., UUVs) with low-altitude platforms (e.g., UAVs). It supports timely network access for UUVs, allowing sustained transmission to facilitate both data return and real-time vehicle control during underwater tasks.
However, the propagation environment for a water-air optical link is exceptionally challenging, as it is affected not only by the two media but also by their interface. At the physics level, optical absorption and scattering by the medium, blockage and beam bending due to debris or other objects, air bubbles in water, turbulence, and refractive-index discontinuities at the interfaces can all distort the optical field. 
From a communication standpoint, these impairments are often abstracted into three dominant classes: received-intensity loss, beam pointing/angular deviation, and arrival delay.
Under these conditions, a water-air optical link is typically implemented as bidirectional optical transceivers, including photoelectricity components (i.e., light-emitting diode and laser diode for transmission as well as photodiode, single-photon avalanche diode and photomultiplier for receiving); optical components (i.e., lens systems and optical filters); pointing/steering devices (i.e., micro-electro-mechanical system and multi-degree-of-freedom  gimbal)
In practice, innovations have been introduced to address the aforementioned issue including received-intensity loss, beam pointing/angular deviation, and arrival delay.
Light with $450\mathrm{nm}$ to $570\mathrm{nm}$ wavelengths, commonly known as "blue and green" light has been widely applied in both scientific research and engineering implementation due to its low absorption and scattering loss in the oceanic environment.
In resistance to the pointing deviation caused by the dynamic ocean interface, beam alignment technology has been applied to adapt angular variation by compensation guided with recursive algorithms like extensive Kalman filters (EKF), or intelligent methods with autonomous sensing of the environment.

\subsection{inter-LAP Network}\label{s2.1.2}
The LAP network serves as the architectural backbone in maritime scenarios, flexibly bridging devices across the lower, peer, and upper layers of the network.
The primary building blocks of the network are LAPs, which maintain links to underwater nodes, provide connectivity to near-space platforms, and perform edge sensing and detection tasks. Therefore, the diversity and complexity of these missions necessitate a multi-LAP network. 
LAP mobility leads to timely variations in physical connectivity, and network topology. Additionally, because airborne platforms typically face limited onboard compute, the LAP layer is often organized in a largely decentralized manner.
In light of the above application context, LAPs in the network are typically equipped with diverse payloads including optical transceivers, vision sensors, and onboard computing components to support communication, sensing, and wireless ad-hoc networking tasks.
%
Network functions often emphasize mobility management, interference coordination, and dynamic topology control to accommodate frequent route changes, spatial variation, and link intermittency. 
%
In a time-varying physical environment, air-to-air links among LAPs experience rapid channel fluctuations, making accurate channel estimation and adaptive modulation essential to mitigate multipath fading and Doppler effects. 
From the perspective of the higher layer, the network’s mobility and flexibility requirements call for joint optimization of the LAP motion and network topology. This coupled problem is highly complex and often intractable for conventional optimization methods, motivating multi-agent and swarm-intelligence approaches for distributed, cooperative controlling, and networking.


\subsection{Near-Space-Air Transmission}\label{s2.1.3}
While LAPs provide deployable capacity for the connection, airships at the near-space layer provide wide-range coverage of the access service, ensuring multiple backhauls from LAPs to the ground station.  
In this situation, airships are required to discriminate LAPs for establishing multi-access networks, which motivates optical integrated communication and sensing (OISAC) and long-range beam alignment.
The links between low-altitude platforms and near-space are often modeled as free-space optical (FSO) communication systems.
In this context, low-altitude-to-space FSO refers to laser links between two layers that provide high-capacity, low-interference backhaul when terrestrial infrastructure is unavailable or out of reach. In practice, overall performance is constrained by the optical link budget, including terminal aperture size, transmit power, and pointing-acquisition-tracking (PAT) stability\cite{UAV_to_Satellite}. 
Additionally, from a channel status perspective, atmospheric effects, including turbulence, scintillation, aerosols, and cloud cover, can cause fading or even temporary outages.
To maintain reliable connectivity under these impairments, such systems typically incorporate channel-aware adaptation such as adaptive modulation and coding, and link-layer reliability mechanisms such as ARQ/HARQ and retransmission strategies.
Compared with RF solutions, FSO links offer much larger usable bandwidth, high directivity with narrow beam divergence, lower probability of intercepting and jamming, and reduced dependence on spectrum availability, making them well-suited for carrying aggregated sensor data, relay traffic, and other time-sensitive mission streams. 

\section{Critical Technologies}\label{s3}
With the development of research in low altitude networking and optical communication, various critical technologies have emerged and are now supporting the network. This section introduces the present development of related technologies in the order of the network structure.


\subsection{Water-Air OWC Channel Modeling}
The complicated cross-media channel is the main influential factor to the water-air optical link. Therefore, accurate channel modeling is a foundational prerequisite and a necessary pathway for further development in the related research.
The challenge of channel modeling arises from the distinct propagation effects in water/air media and the time-varying dynamics of the air-water interface. The combination of these influences leads to a highly intricate channel whose behavior is difficult to capture with a traditional model. Consequently, recent efforts have focused on developing physics-informed channel models that can faithfully characterize the underlying signal propagation\cite{Underwater_and}.
To reconstruct the physical influence on the optical signal, recent research applied the statistical photonic model to approach the channel\cite{Waving_Effect}. 
By modeling the optical beam as an ensemble of photons, the approach represents channel effects stochastically: absorption, scattering, and interface-induced refraction are captured through probability distributions that govern the propagation and angular deviation process of each photon.
Consequently, the calculation results for all photons are aggregated using a Monte-Carlo method that reflects the macroscopic features of the water-air optical channel.
Within this framework, several technical threads are still being developed, including: (i) molecular absorption and scattering models, (ii) underwater turbulence modeling via refractive-index fluctuations, (iii) sea surface dynamics and water-air interface simulation, and (iv) probabilistic characterization of transceiver subsystems. Collectively, these efforts, both ongoing and established, are converging toward a more complete water-air optical channel model. 
Compared to the commonly used exponential-loss abstraction, such physics-informed and probabilistic models offer more precise modeling, guiding both theoretical analysis and system design.



\subsection{Beam Alignment Technology for Water-Air Link}
The unique yet critical challenge of the water-air optical wireless channel is the dynamic ocean surface, which leads to variations in the strength and direction of the optical beam, prompting the need for technologies to compensate for the wave dynamics and maintain a stable communication link between the transceiver\cite{Waving_Effect}.
Beam alignment, also known as beam tracking or beam steering, has emerged as an effective technology to address the aforementioned issue\cite{Underwater_and}. 
%
Beam alignment hardware typically comprises optical lens groups and a mechanical steering unit, and may additionally include sensing and return-path optics such as a quadrant photodiode and a reflector to estimate beam offset. 
Early classical algorithms invoked feedback adjustment based on mean-shift or extended Kalman filtering (EKF), compensating the beam using intensity differences in a quadrant photodiode array and controlling the steering hardware to achieve real-time alignment.
Despite its simple architecture, which makes it accessible, classical algorithms generally require initial pre-alignment and remain passive, with limited ability to sense dynamic sea-surface conditions and compensate proactively.
More recent approaches incorporate learning-based intelligence. One line of work uses deep reinforcement learning (DRL) to adapt  sea-surface dynamics and provide timely pointing corrections. Another leverages proactive vision sensing with convolutional neural networks (CNNs) to extract features and predict near-future channel variations, enabling anticipatory beam steering.
These innovative algorithms achieve high response accuracy and proactive sensing capability; however, they impose higher requirements on platform intelligence and computational support.
So far, designing algorithms that balance latency, accuracy, and resource consumption remains an ongoing optimization target for studies.


\subsection{Swarm Intelligence Enabled Networking}
In LAP swarm applications, the network must operate under harsh conditions: platforms maneuver rapidly, the topology changes continuously, and link quality fluctuates. 
Meanwhile, mission requirements may occur unexpectedly or change abruptly. 
Despite these dynamics, the system is still expected to provide low-latency, reliable connectivity for both command-and-control and payload traffic under tight energy and onboard compute constraints\cite{Swarm}.
Under these conditions, conventional networking designs with static configurations often become unstable: maintaining global topology knowledge is costly, control overhead rises sharply, and the resulting decisions may be delayed behind reality.
Swarm intelligence and distributed learning address this gap by enabling the network to adapt based on experience through local observations and limited neighbor interactions. 
With multi-agent reinforcement learning (MARL), each LAP (or a cluster head) acts as an agent that observes local network and mission cues, including neighbor connectivity, link quality, queue backlog, relative geometry, remaining energy, and chooses actions such as next-hop forwarding, relay selection, congestion-aware scheduling, or transmit power/channel decisions to improve long-term mission performance.
A practical architecture for MARL is centralized training with decentralized execution, where policies are trained on richer data and then deployed as lightweight decision rules that run onboard and react quickly without a fragile central controller. 
%
Graph-structured modeling represents the swarm as a time-varying graph and parameterizes the control policy over local neighborhoods, enabling the learned policy to generalize to swarm scaling and changing topologies.
Alongside MARL, federated learning (FL) and related distributed learning schemes leverage the observation that valid training data are generated throughout the swarm but are challenging to centralize due to bandwidth limits, privacy concerns, and intermittent connectivity. 
In these approaches, LAPs train models locally and share parameter/model updates rather than raw data. Moreover, hierarchical aggregation and asynchronous update protocols can mitigate straggler-induced latency and remain robust under network partitioning.
Compared with conventional approaches, these swarm intelligence and distributed learning designs can reduce control overhead with adaptive broad dissemination, improve robustness through onboard partial information processing, and better handle cross-layer couplings by directly optimizing end-to-end objectives rather than isolated link metrics. 
However, despite these advantages, several key challenges remain. The conflict between learning and communication resource, the non-stationary behavior of the MARL system, the non-IID and drifting data in FL, and the channel limitations imposed by the environment, are all critical issues that need to be further addressed. 
\begin{figure}[h]
    \centering
    \includegraphics[width=1\linewidth]{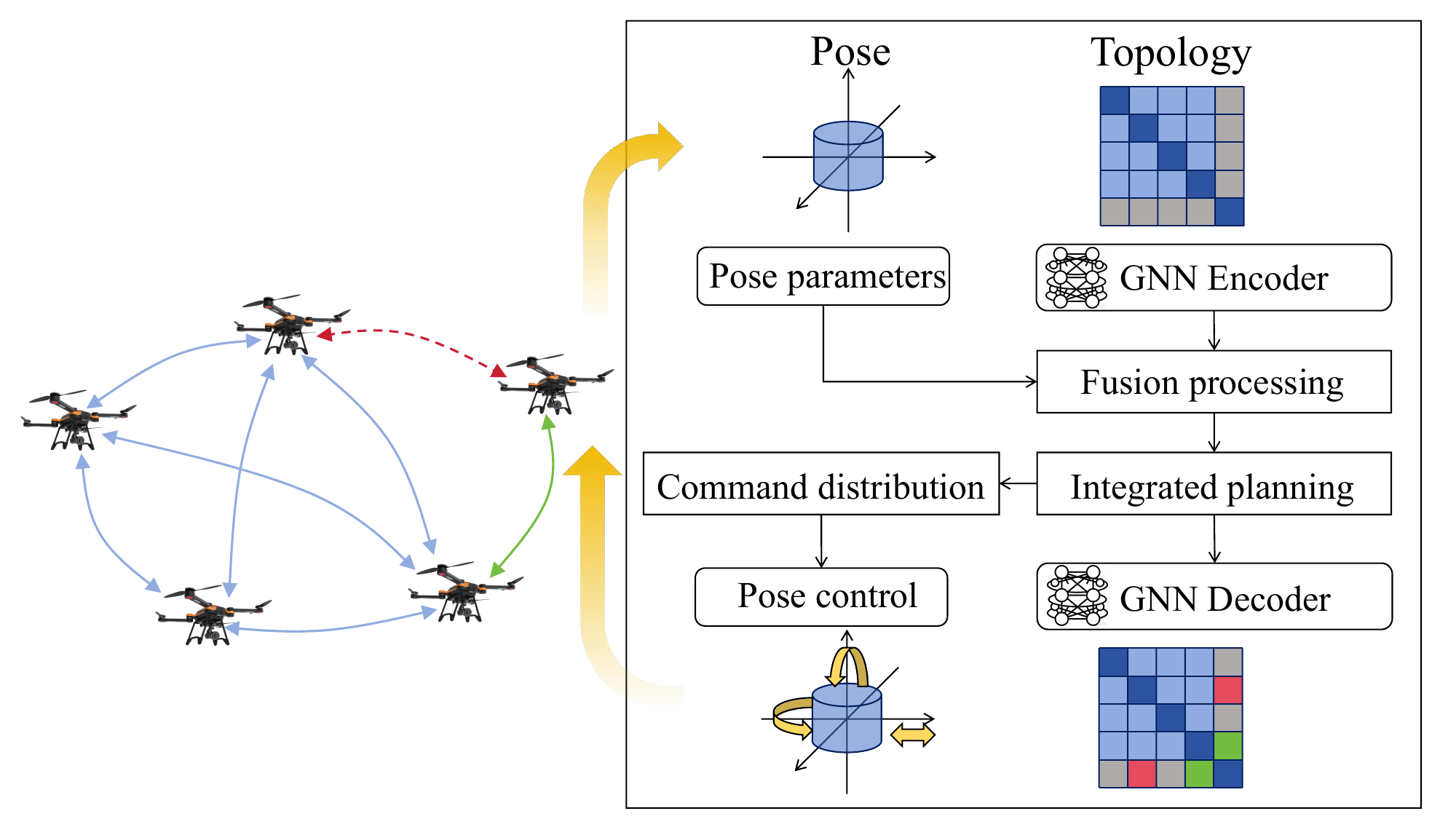}
    \caption{Integrated pose-topology network planning for LAWN.}
    \label{Pose-topology}
\end{figure}

\subsection{Integrated Pose-Topology Network Planning}

Devices in a low-altitude network can be repositioned on demand to match evolving mission objectives. 
Meanwhile, the complexity of cross-layer channels in the maritime environment leads to network topology changes during device mobility\cite{Joint_optimization}.
This makes dynamic, multi-node network planning and online reconfiguration essential to ensure that inter-LAP networks can provide timely and reliable support for oceanic activities.
Previous work has established a technical foundation. 
Trajectory planning algorithms emerge as spatial control methods, which govern LAPs by imposing mathematical constraints, using search algorithms, and employing dynamic programming to obtain a global/local optimal flight trajectory, thereby enabling mobile deployment of LAPs.
In parallel, network planning algorithms address network structure and resource allocation. Through algorithms such as link connectivity management and dynamic bandwidth configuration, the method gets access to network management and resource allocation.
As illustrated in Fig. \ref{Pose-topology}, recent research has attempted to jointly optimize of spatial and network planning, including trajectory/flight-attitude optimization under signal strength constraints, and network design with positional-weighted resource allocation. 
However, most of the aforementioned approaches have been developed and validated primarily in simplified settings and rely on strong modeling assumptions. As a result, they often fall short of properly describing and effectively guiding comprehensive systems in marine environments.
Integrated pose-topology network planning is an innovative technical paradigm that jointly incorporates spatial information from device platforms and network information into a single model. 
By establishing a relationship between physical states and network topology via the environment-channel relationship, it further jointly optimizes device-state control and network resource allocation based on the mission.
This method involves multiple influencing factors of device control, optical channel, and network management, making it suitable for applying deep learning algorithms to achieve end-to-end joint optimization, or using reinforcement learning to learn and react to environmental parameters adaptively.

\begin{figure}
    \centering
    \includegraphics[width=1\linewidth]{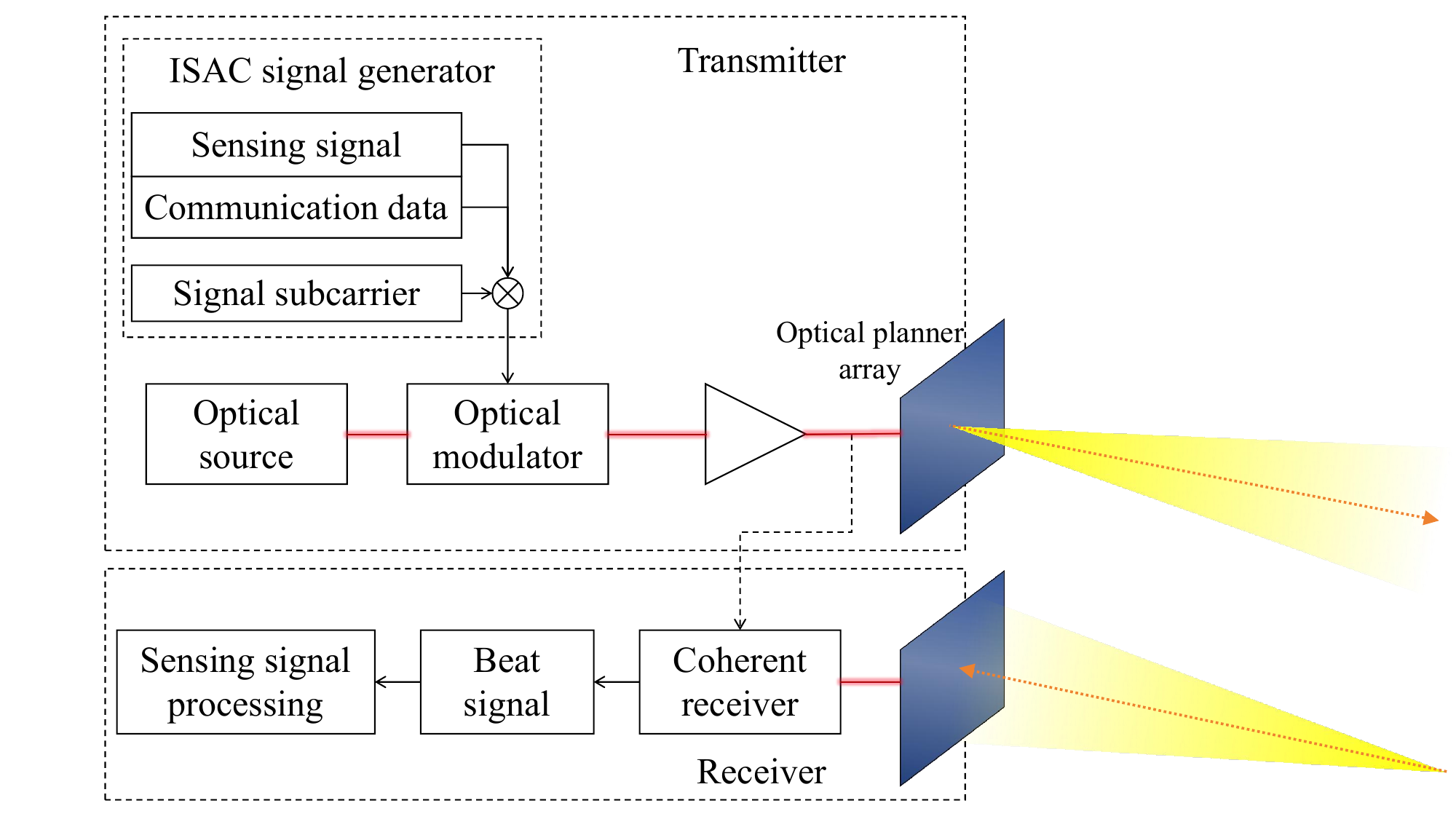}
    \caption{The architecture of an O-ISAC system.}
    \label{OISAC}
\end{figure}

\subsection{Free-Space Optical Integrated Sensing and Communication}
The requirements of linking Near-Space Platforms with LAPs have driven the implementation of Optical Integrated Sensing and Communication (O-ISAC)\cite{DCO-OFDM}. 
These links require simultaneous high-capacity data transfer and high-fidelity environmental perception. Traditional radio frequency (RF) approaches face inherent limitations due to spectral scarcity and limited resolution, posing a significant challenge for achieving efficient and unified communication-sensing capabilities in dynamic aerial environments without performance trade-offs\cite{ISAC_LAE}.
As shown in Fig. \ref{OISAC}, O-ISAC offers a transformative solution by leveraging the vast bandwidth and ultrashort wavelength of optical carriers to enable simultaneous high-resolution sensing and high-capacity communication over a single optical link. 
The transmitted optical carrier enables multi-gigabit data transmission while also allowing precise derivation of LAP kinematic states and nearby environmental features through advanced analysis of retroreflected and scattered signals. Furthermore, this dual functionality addresses the synergistic demands of connectivity and situational awareness. It not only achieves exceptional spectral efficiency but also enhances system security through spatial confinement and a low probability of intercept. 
Thus, O-ISAC provides a significant foundation for spectrally efficient, secure, and autonomously coordinated operations across cross-layer aerial links.
While promising, the practical deployment of O-ISAC for linking NSPs with low-altitude LAPs must address several key considerations. 
These are rooted in complex atmospheric dynamics, current hardware constraints, and the need for a unified theoretical framework to guide the deep synergy between communication and sensing. Advancing this technology involves developing intelligent, adaptive systems capable of dynamic waveform design and predictive beam steering guided by real-time environmental perception. 
Furthermore, innovations in hybrid photo-electronic hardware are crucial to realize compact multifunctional transceivers. 
By developing joint communication-sensing waveforms tailored for optical systems, supported by cross-layer optimization and machine learning, O-ISAC is poised to become the cornerstone of secure, efficient, and self-aware hierarchical aerial networks.

\begin{figure*}
    \centering
    \includegraphics[width=1\linewidth]{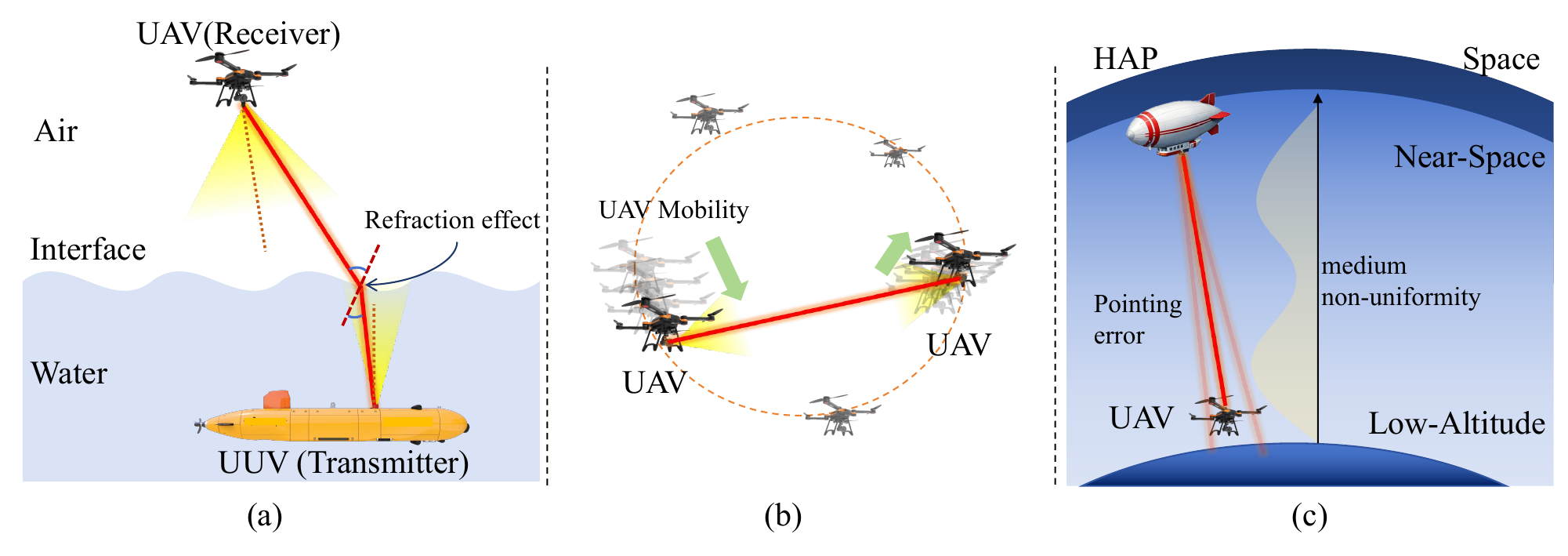}
    \caption{Optical wireless channel model for different communication links in the optical LAWN system. (a) Water-to-air direct optical link. (b) Optical wireless link between dynamic low-altitude platforms. (c) FSO link between low-altitude and near-space.}
    \label{channel}
\end{figure*}

\section{Future Research and Open Challenges}\label{s4}
\subsection{Optical Wireless Channel Modeling}
%
In the network, OWC is introduced to support various link types, including low-altitude-near-space links, air-to-air links, and water-to-air cross-medium links. 
Meanwhile, due to markedly higher carrier frequencies, optical waves exhibit fundamentally different propagation behavior than RF waves. Accordingly, RF channel models are not directly applicable to optical wireless channels without substantial modification.
These situations necessitate precise and faithful optical wireless channel models which effectively describe the propagation characteristics of the optical signal, serving as a foundation for further research on OWC systems.
Because propagation conditions vary widely across use cases, channel modeling must be tailored to the target scenario. Accordingly, the channel can be categorized into the following link configurations:
\subsubsection{Water-Air OWC Channel}
As illustrated in Fig. \ref{channel} (a), the main issue of the water-to-air OWC channel is the two distinct media through which the optical beam traverses. In particular, surface dynamics randomly reshape the local incidence geometry, thereby inducing rapid fluctuations in both received power and beam pointing.
Recent work has already approached the problem from various aspects. On the surface modeling side, spectrum-driven harmonic models can synthesize ocean waves, but such models are often idealized, failing to capture without capturing breaking/plunging waves and the bubble fields they generate\cite{Waving_Effect}. Therefore, a wave model that can effectively account for boundary roughness and scattering in practice represents a key direction for future research. 
On the propagation modeling side, recent work has introduced geometrical-optics models and photon Monte-Carlo statistical simulators. However, limited by simplifying assumptions and computational cost, these simulators typically match controlled laboratory conditions and do not reliably extrapolate to operational environments. This motivates the development of faithful cross-interface models, together with computationally efficient simulation pipelines. 
Notably, a few recent studies have explored light-field formulations to establish OWC models and further introduced structured light to address channel impairments. Although previous have considered only underwater scenarios, these methodologies may be extended, likely pointing to another promising direction.
\subsubsection{inter-Low-Altitude Channel}
As illustrated in Fig. \ref{channel} (b), the core issue of low-altitude LAP-to-LAP channels is the mobility of both communication ends, which causes frequent pointing variations and Doppler effects. 
Research in this direction remains nascent and mostly focuses on simplified system modeling for bistatic or monostatic scenarios, while lacking studies on communication channels between two dynamically moving terminals and methods to cope with Doppler effects. 
As low-altitude aerial networks evolve, dynamic air-to-air channel modeling and Doppler estimation/compensation will become essential research priorities.
\subsubsection{FSO Channel Between Low-Altitude and Near-Space}
As illustrated in Fig. \ref{channel} (c), the core issues of FSO channels between low-altitude and near-space are the medium nonuniformity of the channel and the amplification of pointing errors in long-distance communications. 
A low-altitude-to-near-space channel spans multiple meteorological layers with different characteristics, leading to the signal attenuation and scattering due to medium nonuniformity and significant atmospheric turbulence\cite{FSO_backhaul}. 
Therefore, establishing a non-uniform channel model span multiple meteorological layers is a future research directions. 
On the other hand, the link distance between low-altitude and near-space platforms can reach 10-100 km, causing even minor angular misalignment to translate into a large displacement. Therefore, accuracy becomes the prior consideration of the model.

In conclusion, optical wireless channel modeling will be a complex but significant topic, and it will remain a long-term research issue.


\subsection{Precise Optical Beam Alignment}
Across various operating scenarios, links between low-altitude and near-space exhibit extreme sensitivity to pointing errors at long ranges, whereas water-to-air channels vary rapidly as sea-surface dynamics continuously perturb the effective interface. 
Conventional schemes, which are often built around a four-quadrant photodiode (4QPD) and PID feedback control, struggle under fast fading, abrupt disturbances, and highly nonstationary conditions\cite{Underwater_and}. 
Though recent studies have proposed more advanced methods through algorithm compensation, specific-task optimization, and intelligent augmentation, the need for high-sensitivity, environment-adaptive beam alignment and tracking technologies remains.
One possible direction is sense-driven adaptive alignment, in which the optical link is augmented with sensing techniques such as optical ISAC, photosensor arrays, or vision-based feedback. 
By sensing and estimating channel and environmental states, combined with prior channel models, this optical beam alignment scheme can enable robust closed-loop control for stable beam tracking and sustained alignment.


\subsection{Solar Noise Mitigation}
Solar noise is a common challenge that influences every single equipment exposed to sunlight in open environment OWC systems. 
As sunlight carries substantial energy across a broad spectral range, it cannot be entirely suppressed by optical filtering and affects photosensitive sensors at the receiver through direct illumination, atmospheric scattering, and sea-surface reflection. High-intensity solar noise, such as direct or reflected sunlight, can saturate photosensitive sensors, blocking signal transmission\cite{Background_Noise}. 
Thus it is necessary to avoid direct/reflected solar paths by beam steering. In addition, atmospheric scattering noise is widespread and requires signal-processing algorithms to separate it from signals. 
Moreover, solar noise exhibits an asymmetry between uplink and downlink. Due to differences in orientation and medium attenuation, the transmitter and receiver experience different levels and statistics of solar noise. Therefore, uplink and downlink cannot simply reuse configuration parameters based on the principle of optical path reversibility, and direction-aware adaptive compensation against solar noise is needed.
In conclusion, solar-noise mitigation emerges as a pervasive, system-level challenge. 
Its realization and optimization typically require tight co-operation across optical components, channel modeling, environment sensing, and signal-processing algorithms, and it can significantly impact the performance of all links in the system. 
Accordingly, solar-noise mitigation is a critical direction for optical LAWN.

\subsection{Agentic Intelligence for Physical Layer}
As communication systems grow more complex and low-altitude edge devices become more capable, local agentic intelligence is no longer just feasible, but increasingly necessary. 
This trend calls for physical-layer intelligence at the edge to enable channel-state awareness, modulation, and waveform adaptation, and closed-loop cooperative control.
In particular, the bottleneck of physical-layer intelligence is less about peak compute throughput than about real-time response and strict correctness under tight latency and reliability constraints. 
This attracts attention toward on-chip neural processing, efficient inference pipelines, and application-aware intelligence tailored to specific tasks. 
In addition, to interface with higher-layer applications and swarm intelligence, cross-layer interfaces across protocol layers are a key development priority.
In conclusion, agentic intelligence for physical-layer forms a next-generation foundation of the optical LAWN network, yet there are still many directions to explore.

\section{Conclusion}
This article provides an overview of the optical low-altitude network from a systematic perspective and analyzes technologies and open challenges of the optical wireless network structure across near-space, low-altitude, ground, and sea scenarios. 
Due to the unique oceanic scenario, optical wireless communication will be a promising method for low-altitude networks, offering high-rate transmission or cross-media communication capabilities. 
However, realizing such a system remains challenging because of the complex channel environment. 
To address these issues, a range of techniques, including O-ISAC, beam alignment, autonomous networking, and joint pose-topology optimization, are being developed and applied from the physical layer up to the network layer.
Additionally, the development of artificial intelligence and agentic terminal devices has further enlarged the superiority of such network architectures.
Yet there are still various issues to be addressed, including channel modeling, beam alignment, and environmental influence compensation. This, in turn, highlights the potential for the system to improve and shows the existing opportunities for further research and exploration in this area.

\bibliography{Reference.bib}
\bibliographystyle{IEEEtran}

\vfill

\end{document}